\documentclass{mem}
\usepackage{natbib}\usepackage{txfonts}\usepackage{balance}
\usepackage{graphicx}
\usepackage[a4paper,breaklinks,dvipdfm]{hyperref}
\idline{84}{1}
\usepackage{txfonts} 

\begin{document}
\def\teff{$T\rm_{eff }$}
\def\kms{$\mathrm {km s}^{-1}$}

\title{
High redshift blazars
}

   \subtitle{}

\author{
G. Ghisellini\inst{1} 
          }

  \offprints{G. Ghisellini}

\institute{
INAF --
Osservatorio Astronomico di Brera, Via Bianchi 46, I--23807 Merate (Lecco), Italy
\email{gabriele.ghisellini@brera.inaf.it}
}

\authorrunning{Ghisellini }

\titlerunning{High redshift blazars}

\abstract{Blazars are sources whose jet is pointing to us.
Since their jets are relativistic, the flux is greatly amplified in the
direction of motion, making blazars the most powerful persistent
objects in the Universe. This is true at all frequencies, but especially
where their spectrum peaks.
Although the spectrum of moderate powerful sources peaks in the $\sim$GeV range,
extremely powerful sources at high redshifts peak in the $\sim$MeV band.
This implies that the hard X--ray band is the optimal one to find 
powerful blazars beyond a redshift of $\sim$4.
First indications strongly suggest that powerful high--$z$ blazars
harbor the most massive and active early black holes, exceeding a billion solar masses.
Since for each detected blazars there must exist hundreds of similar, but
misaligned, sources, the search for high--$z$ blazars is becoming competitive with the
search of early massive black holes using radio--quiet quasars. 
Finding how the two populations of black holes (one in jetted sources, the other in radio--quiet 
objects) evolve in redshift will shed light on the growth of the most massive black holes and
possibly on the feedback between the central engine and the rest of the host galaxy.
}
\maketitle{}

\section{Introduction}

Blazars are jetted Active Galactic Nuclei (AGN), whose jet is
pointing close to our line of sight.
How close?  A convenient definition is to require that the
viewing angle $\theta_{\rm v}<1/\Gamma$, where $\Gamma$ is the
bulk Lorentz factor of the jet emitting plasma.
With this definition for each detected blazars there must exist
other $2\Gamma^2=200(\Gamma/10)^2$ misaligned sources sharing the same 
intrinsic properties, but appearing dramatically different.
The enhancement due to relativistic beaming make blazars 
visible even at high redshifts, and this makes them useful
probes to explore the far Universe, even if radio--quiet
quasars are $\sim$10 times more numerous than radio--loud sources,
and then $\sim 10^3$ times more numerous than blazars.
Although high--$z$ blazars might be rare, their
detection require the existence of a much larger population of
similar sources.
In particular, if the black hole mass of a blazar at $z=5$ is
larger than a billion solar masses, there must exist hundreds of
equally heavy black holes in the misaligned sources.
This motivates our search for high--$z$ blazars, since
it is competitive with the analogous search of heavy black holes
in high--$z$ radio--quiet quasars.

There are hints that radio--loud sources are associated to 
larger black hole masses (e.g. Laor 2000; Chiaberge \& Marconi 2011, 
but see Woo \& Urry 2002) and to very massive
elliptical hosts (with the possible exception of radio loud Narrow Line Seyfert 1
galaxies, see Foschini et al. 2011). 
This makes us wonder if this is true also at large redshifts (i.e. $z>3-4$).
Is the presence of a relativistic jet associated to a black
hole of a larger size even when the first supermassive black holes
were formed?

\section{The SED of powerful high--$z$ blazars}

\begin{figure}[]
\vskip -0.3 cm
\resizebox{\hsize}{!}{\includegraphics[clip=true]{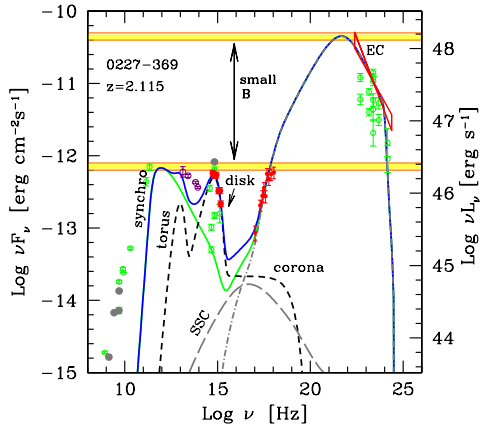}}
\vskip -0.5 cm
\caption{
\footnotesize
The SED of PKS 0227--369, a $\gamma$--ray bright blazar at $z=2.115$, 
detected by the {\it Fermi} satellite. 
Data (references in Ghisellini et al. 2009) 
have been updated with the {\it WISE} IR fluxes (purple points).
The solid blue line is the result of applying a one--zone leptonic model
(Ghisellini \& Tavecchio 2009): the different components are labelled.
The large Compton to synchrotron ratio (see the horizontal bars),
that is common in these powerful sources, set limits
on the strength of the magnetic field. 
The accretion disk emission is well visible.
}
\label{0227}
\end{figure}

Fig. \ref{0227} shows the SED of PKS 0227--369 ($z=2.115$), a typical powerful 
and $\gamma$--ray bright blazar.
Detected by the {\it Fermi} satellite, its $\gamma$--ray luminosity
exceeds $10^{48}$ erg s$^{-1}$.
The $\gamma$--ray component dominates over the synchrotron luminosity by almost two
orders of magnitude.
Therefore the magnetic field in the emitting region cannot be too large,
and this is the reason why the jet cannot be magnetically dominated at
the jet scales where most of the luminosity is produced 
(e.g. Celotti \& Ghisellini 2008).
The peak of the synchrotron emission lies in the mm range, and the steep
synchrotron spectrum beyond the peak let the accretion disk flux
to dominate the optical--UV spectrum.
Applying a simple Shakura \& Sunyaev (1973) disk emission model 
we can fit the contribution of the disk to the SED, and find the
black hole mass $M$ and the disk luminosity $L_{\rm d}$.
In addition, we can estimate $L_{\rm d}$ using the broad emission lines
(these powerful blazars are all Flat Spectrum Radio Quasars, with
strong broad optical emission lines), especially when the optical
continuum is contaminated by some 
non--thermal synchrotron flux.
Details of the method can be found in Calderone et al. (2012).

\subsection{Hard X--rays are better than $\gamma$--rays} 

\begin{figure}[]
\vskip -0.3 cm
\resizebox{\hsize}{!}{\includegraphics[clip=true]{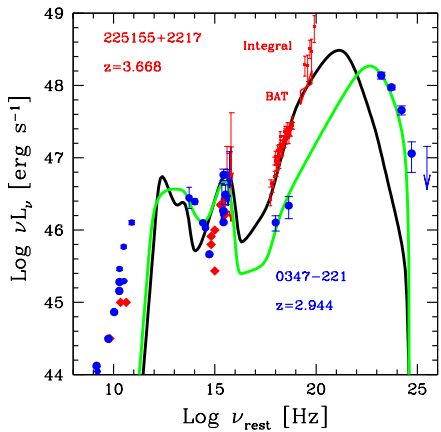}}
\vskip -0.5 cm
\caption{
\footnotesize
The SED of
PKS 0347--211 compared to
MG3 225155+2217 (Ghisellini et al. 2009), to show that the most powerful blazars
at high redshifts are relatively weaker $\gamma$--ray and stronger hard X--ray
sources. This is due to the shift to smaller frequencies of the emission peaks
(according to the so called {\it blazar sequence,} Fossati et al. 1998).
To this  intrinsic effect we must add the impact of the K--correction,
that favors the hard X--ray band of sources at a larger redshift.
}
\label{225155}
\end{figure}

Fig. \ref{225155} shows the SED of two powerful blazars, at a
different redshift: MG3 215155+2217 ($z$=3.668) is a strong hard X-ray source,
detected by the BAT instrument [15--150 keV] onboard the {\it Swift} satellite,
but not by {\it Fermi}.
On the contrary PKS 0347--221 is a very powerful {\it Fermi} source,
but weaker in the X--ray band, and not detected by BAT.
This is due to the different location of the high energy peak,
being at lower frequencies in MG3 215155+2217, the more powerful
source. 
This in agreement with the {\it phenomenological blazar sequence} (Fossati et al. 1998).
Furthermore, for increasing redshift, the K--correction favors the detection 
of the hard X--ray flux and works against the detection of the steep $\gamma$--ray 
component.

It is for these two reasons that the best way to find high--$z$ blazars is through
observations in the hard X--ray band, close to where the high energy component peaks
in very powerful sources.
It is not by chance that BAT detects blazars at a larger redshift than {\it Fermi}/LAT
(see Ajello et al. 2009; 2012).
Therefore the best way to find high redshift blazars would be a 
survey in the hard X--rays (above 10 keV), as was envisaged by the {\it EXIST} mission
(Grindlay et al. 2010; see also Ghisellini et al. 2010a for the number of detectable 
high--$z$ blazars).
In the absence of an hard X--ray survey, we must find good 
blazar {\it candidates} on the basis of observations coming from the radio and optical
(giving us the radio--loudness and the redshift) and from the X--rays (flux and slope).
Then we should confirm the blazar nature through dedicated 
observations in the hard X--ray band.

Why is it so crucial to observe at high X--ray energies?
Because at these frequencies we are close to the emission peak of the jet emission:
a large hard X--ray to optical ratio indicates a highly beamed source.
The same indication comes from the radio--loudness, if the optical is dominated
by the unbeamed accretion disk flux, and the radio by the beamed jet
(in principle, at 1.4 GHz there might be some contribution from the unbeamed radio--lobe,
even if it is unlikely for high--$z$ sources).
The added value of the X--ray data (and their crucial role in classifying the source as a blazar)
comes because the main radiation process at these frequencies is the inverse Compton scattering 
by relativistic electrons in the jet and the
seed photons produced in the broad line region (BLR) and,
possibly, the IR photons from the dusty torus (this process is called {\it External} Compton,
or EC for short).
In the comoving frame, the seed photons are highly anisotropic, making the scattered flux
anisotropic even in the comoving frame (Dermer 1995).
In the observer frame, this translates in a beaming pattern different
than the one of the synchrotron flux (it is more enhanced in the forward direction and
more depressed at large angles).
Therefore, for decreasing viewing angles, the X--ray (EC) flux increases more than
the radio (synchrotron) flux.
Furthermore, the [0.3--10 keV] X--ray EC spectrum is predicted to be very hard, 
even harder than $\alpha_X\sim 0.5$, where $\alpha_X$ is defined through 
$F(\nu)\propto \nu^{-\alpha_X}$.

\section{Chasing high redshift blazars} 

The record holder of the largest redshift for blazars
is Q0906+6930, with $z=5.47$ (Romani et al. 2004).
There is a possible association of this object with an EGRET
source (but the $\gamma$--ray detection is only at the $\sim$3$\sigma$ significance level), 
while {\it Fermi}/LAT did not detect it (yet).
Romani (2006) estimated a virial black hole mass $M\sim 2\times 10^9 M_\odot$.

The discovery of this blazars came serendipitously, not through 
the study of a well defined sample, in terms of flux limit and covered sky area.
Since the comoving volume between $z=5$ and $z=6$ is 380 Gpc$^3$,
its existence implies the existence of more than $450(\Gamma/15)^2/380\sim 1$ Gpc$^{-3}$
black holes with mass $M>10^9 M_\odot$ in this redshift bin
(see Fig. 16 in Ghisellini et al. 2010a).
This is not much smaller than the number density 
of radio--quiet quasars of optical luminosity larger than $10^{47}$ erg s$^{-1}$
(Hopkins et al. 2007), that must host a black hole heavier than a billion solar masses
not to become super--Eddington
(see Volonteri et al. 2011 for a discussion about the radio--loud fraction at these 
redshifts and the problem of the paucity of radio--loud sources in the SDSS+FIRST survey).

Ajello et al. (2009) studied the blazars detected by the {\it Swift}/BAT instrument,
restricting the energy range to [15--55 keV] to have a cleaner signal.
They also derived the luminosity function of these hard X--ray blazars, and
noted that the number density of the most powerful objects 
[above $\log(L_X /{\rm erg\, s^{-1}}) =47.2$]
has a well pronounced peak around $z\sim4$.
We (Ghisellini et al. 2010a) have modeled these powerful objects and 
found that all have a black hole with $M>10^9 M_\odot$, accreting 
on average at the $\sim$10\% Eddington rate.
The relatively large $L_{\rm d}/L_{\rm Edd}$ ratio tells us that we are observing 
the end of the formation process of a heavy black hole.

We then started another approach to find high--$z$ blazar candidates in a 
systematic way.
We selected all SDSS quasars (Shen et al. 2011) that are also detected by the FIRST survey
(at 1.4 GHz, with a flux limit around 1 mJy) above $z=4$.
To select sources with a small viewing angle, we also
required the radio--loudness $R$ to be larger than 100
[$R\equiv F(5 \, {\rm GHz})/F(2500 \AA) $].
We find 31 objects, including 3 objects at $z>5$.
Of these 31 sources, 19 were observable from La Silla (Chile) with the
GROND instrument (Greiner et al. 2008), able to observe simultaneously in
3 IR and 4 optical filters.
Together with the far IR data from {\it WISE}, we have a very good
coverage of the IR--optical spectrum, where we expect the peak of the the accretion disk
component (at these redshifts).
The results of this study are in preparation (Sbarrato et al. 2013),
but during this work we realized that the most distant source of the
sample (at $z\simeq 5.3$) was also the best candidate to be a blazar
with a large black hole mass. 
This source is described below.

\begin{figure}[]
\resizebox{\hsize}{!}{\includegraphics[clip=true]{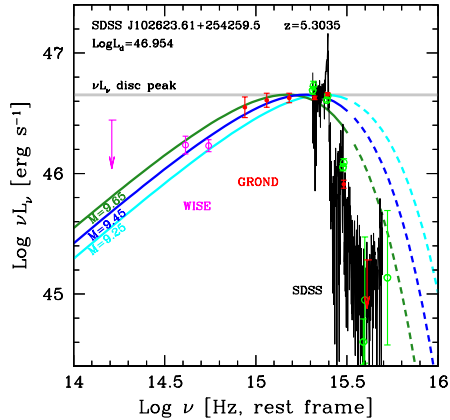}}
\vskip -0.5 cm
\caption{\footnotesize
The IR to optical SED of B2 1023+25.
Data are from {\it WISE} (purple) GROND (red) and SDSS (black line,
absorbed blueward of the the Ly$\alpha$ line.
We show three Shakura \& Sunyaev disk spectra, with the same $L_{\rm d}$ 
but slightly different masses (whose logarithmic value is labelled), 
to appreciate the uncertainties related to the mass estimate. From 
Sbarrato et al. (2012).
}
\label{mass1023}
\end{figure}

\subsection{The second most distant blazar}

Sbarrato et al. (2012) realized that B2 1023+25 (at $z=5.3$)
was a very good blazar candidate, 
on the basis of its very large radio--loudness ($R\sim 5,000$) 
and radio flux (261 and 106 mJy at 1.4 and 8.4 GHz, respectively).
Fig. \ref{mass1023} shows the IR--optical SED:
thanks to the photometric {\it WISE} and GROND
observations and partially by the SDSS spectrum [that is however absorbed 
blueward of the Ly$\alpha$ ($\log \nu \sim 15.4$)], we could determine, 
in a reliable way, the black hole mass: $M\sim 3\times 10^9M_\odot$, 
with an uncertainty of less than a factor 2. 
We also found that the disk luminosity is $\sim$1/4 Eddington.

\begin{figure}[]
\vskip -0.3 cm
\resizebox{\hsize}{!}{\includegraphics[clip=true]{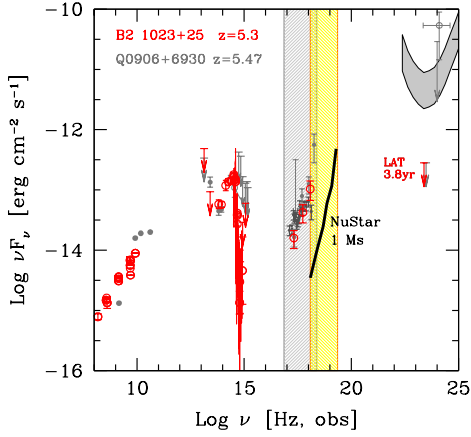}}
\vskip -0.5 cm
\caption{\footnotesize
The entire SED of B2 1023+25 (larger red empty circles)
compared to the SED of Q0906+6930 (smaller grey filled circles).
Note that they are almost identical, apart from the EGRET 
(uncertain) detection. See also the upper limit given by {\it Fermi}/LAT.
We show the limiting flux of {\it NuStar} for 1 Ms exposure time. Both
sources could be detected by {\it NuStar} in $\sim$$10^4$ s. From 
Sbarrato et al. (2012).
}
\label{1023}
\end{figure}

Then we requested a ToO observation from the {\it Swift} team, in order to 
confirm our hypothesis. 
The total exposure of the {\it Swift}/XRT instrument [0.3--10 keV]
was $\sim$10 ks and resulted in 26 counts. 
Despite the small number of detected photons, the X--ray flux and slope was 
entirely consistent with the blazar hypothesis.
Fig. \ref{1023} shows the entire SED of the source, and compares it
with the SED of Q0906+6930. 
The two SEDs are almost identical. 
The X--ray flux is strong and hard: even in the relatively soft
{\it Swift}/XRT band  
the X--ray flux exceeds the accretion disk flux.
The same figure shows the limiting (differential) flux of {\it NuStar} for an exposure of 1 Ms.
It is clear that both B2 1023+25 and Q0906+6930 can be detected by {\it NuStar}
up to 80 keV with a very short exposure.
If the flux keeps to be hard up to a few tens of keV, this would be the final proof 
of the blazar nature of both objects.

This is important, since only if these two sources are blazars (namely, if their $\theta_{\rm v}<1/\Gamma$),
we can can calculate how many similar but misaligned sources there are in this redshift bin.
Assuming that B2 1023+25 is indeed viewed at $\theta_{\rm v}<1/\Gamma \sim 3.8^\circ(15/\Gamma)$,
then there must be other $\sim 450 (\Gamma/15)^2$ sources with a black hole mass
similar to B2 1023+25, with the similar redshift, and in the portion of the sky covered
by the SDSS+FIRST survey.
The latter covers 8,770 square degrees.
Therefore, in the redshift bin $5<z<6$, the existence of B2 1023+25
implies the existence of other
$450(\Gamma/15)^2 (40,0000/8,770)\sim 2,000$ jetted sources
in the same redshift bin and all sky, all having a black hole mass $M>10^9 M_\odot$.
Dividing for the comoving volume (i.e. 380 Gpc$^3$)
we have a density $\phi(M>10^9M_\odot)$=5.4  Gpc$^{-3}$ in the $5<z<6$ redshift bin.
Note that this is a lower limit, since other high redshift blazars could exist, 
but not yet identified as such (the existence of Q0906+6930 does not change
this estimate, since it is one of the blazars expected in the portion of the sky
not covered by the SDSS+FIRST survey).

\section{Early supermassive black holes}

At a redshift $z=5.3$, the age of the Universe is 1 billion years.
This is a strict upper limit for the time needed to build up
a black hole of mass equal to $3\times 10^9 M_\odot$.
If the black hole accretes at the rate $\dot M$, and produces a luminosity
$L_{\rm d}= \eta \dot M c^2$, the time needed to double the 
black hole mass is the Salpeter time $t_{\rm S}$:
\begin{equation}
t_{\rm S} \, = \,  {\eta \sigma_{\rm T} c \over 4\pi Gm_{\rm p}} \, {L_{\rm Edd}\over L_{\rm d}} 
\sim \, 45\,  \eta_{-1} {L_{\rm Edd}\over L_{\rm d}}\,\, {\rm Myr}
\end{equation}
where $\eta=10^{-1}\eta_{-1}$.
Assuming a black hole seed of mass $M_0$, and that the accretion process
occurs always with the same efficiency $\eta$, the time $\Delta t$ needed to
build up the final mass $M$ is
\begin{equation}
\Delta t \, =\, t_{\rm S} \, \ln \left( {M\over M_0} \right)
\end{equation}
If the accretion occurs at the Eddington rate all the time, with $\eta=10^{-1}$,
the time needed to reach $3\times 10^9 M_{\odot}$ is 0.77 or 0.46 Gyr for a seed mass
of 100 or $10^5$ solar masses, respectively.
If we find a similar black hole mass at $z=7$, when the Universe was 0.75 Gyr old, 
then we can exclude that the black hole seed had the mass of typical 
PopIII star remnant black holes, or else that the accretion was 
always near or sub--Eddington, or that the efficiency was as large as 10\%.

Indeed, Mortlock et al. (2011) found a radio--quiet quasars, ULAS J1120+0641, at $z=7.085$
hosting a black hole with a mass $M\sim 2\times 10^9 M_\odot$.
This was estimated through the virial method (i.e. FWHM of the MgII 
broad line and the optical continuum luminosity).
If true, this high value for the black hole mass is difficult
to reconcile with the simple scenario of black hole
growth through accretion at the Eddington rate and with $\eta\sim$0.1.

\section{Conclusions}

The rationale for using blazars as probes of the far Universe is
based on the fact that they appear very powerful, and can be 
detected at very large redshifts.
While in the optical they are identical to radio--quiet
quasars, their X--ray emission stands out, especially
at large energies (tens of keV), since their X--ray spectrum is hard.
Therefore hard X--ray surveys would be the best way to find them out.
In these objects we can observe at the same time the non--thermal 
jet emission and the accretion disk component, left ``naked" (i.e. not
covered) by the synchrotron flux that peaks in the mm band and has a steep
spectrum after the peak.

Besides their cosmological use, in these sources we can directly study
how the power of the jet is associated to the accretion rate.
A strong correlation is present (Ghisellini et al. 2010b; Sbarrato et al. 2012),
with the jet power that is often larger than the disk luminosity.

The finding that jetted AGNs are on average hosting black holes of larger 
mass than radio--quiet ones is intriguing, albeit controversial,
and we hope to investigate this very issue at redshifts larger than 4.
To this aim, while awaiting for hard X--ray mission capable of performing
surveys deeper than BAT or INTEGRAL, we must rely on large radio and optical
surveys (i.e. SDSS+FIRST) and be able to select the best high--$z$ blazar candidates,
to be confirmed by dedicated X--ray observations, preferentially in the hard band,
above 10 keV.


\bibliographystyle{aa}

\end{document}